\documentclass[10pt]{article}

\usepackage{amsmath}
\usepackage{amssymb}

\usepackage{graphicx}

\usepackage{cite}

\usepackage{color}

\usepackage{soul}

\usepackage[labelfont=bf,labelsep=period,justification=raggedright]{caption}

\makeatletter
\renewcommand{\@biblabel}[1]{\quad#1.}
\makeatother

\date{}

\begin{document}

\begin{flushleft}
{\Large
\textbf{Transfer Entropy reconstruction and labeling of neuronal connections from simulated calcium imaging}
}
\\
Javier G. Orlandi$^{1}$,
Olav Stetter$^{2,3,4}$,
Jordi Soriano$^{1}$,
Theo Geisel$^{2,3,4}$,
Demian Battaglia$^{2,4,5\ast}$
\\
\textbf{1} Departament d'Estructura i Consituents de la Mat\`eria, Universitat de Barcelona, Barcelona, Spain
\\
\textbf{2} Max Planck Institute for Dynamics and Self-Organization, G\"ottingen, Germany
\\
\textbf{3} Georg-August-Universit\"at, Physics Department, G\"ottingen, Germany
\\
\textbf{4} Bernstein Center for Computational Neuroscience, G\"ottingen, Germany
\\
{\textbf{5} Institut de Neurosciences des Syst\`emes, Inserm UMR1106, Aix-Marseille Universit\'e, Marseille, France}
$\ast$ E-mail: {demian.battaglia@univ-amu.fr}
\end{flushleft}

\section*{Abstract}

Neuronal dynamics are fundamentally constrained by the underlying
structural network architecture, yet much of the details of this synaptic
connectivity are still unknown even in neuronal cultures \emph{in vitro}.
Here we extend a previous approach based on information theory, the
Generalized Transfer Entropy, to the reconstruction of connectivity of simulated neuronal networks of
both excitatory and inhibitory neurons.  We show that, due to the model-free
nature of the developed measure, both kinds of connections can be reliably
inferred if the average firing rate between synchronous burst events exceeds a
small minimum frequency. Furthermore, we suggest, based on systematic
simulations, that even lower spontaneous inter-burst rates could be raised to
meet the requirements of our reconstruction algorithm by applying a weak
spatially homogeneous stimulation to the entire network. By combining
multiple recordings of the same \emph{in silico} network before and after
pharmacologically blocking inhibitory synaptic transmission, we show then how it
becomes possible to infer with high confidence the excitatory or inhibitory nature of
each individual neuron.

\section*{Introduction}

Important advances in the last decade have provided
unprecedented detail on the structure and function of brain
circuits~\cite{Bullmore:2009iv,Power2011665} and even programs aiming at an
exhaustive mapping of the brain connectome(s) have been announced
\cite{HumanBrainProject,BigProjects, BrainActivityMap1,BrainActivityMap2}.
First, the combination of invasive and non-invasive techniques such as
high-resolution optical imagery and diffusion-based tractography have
revealed the major architectural traits of brain circuitry
\cite{Hagmann:2008gd}. Second, functional imaging has
provided non-invasive measures of brain activity, both at
rest~\cite{Deco2011Emerging} and during the realization of specific
tasks~\cite{Power2011665}. These efforts have opened new perspectives in
neuroscience and psychiatry, for instance to identify general principles
underlying interactions between multi-scale brain circuits \cite{Varela:2001bs,
Raizada:2003wd}, to probe the implementation of complex cognitive processes
\cite{Corbetta:2008ea, Gaillard:2009fd}, and to design novel clinical prognosis
tools by linking brain pathologies with specific alterations of connectivity and
function \cite{Zhou2012Predicting, Seeley2009Neurodegenerative,Mary-Ellen2010Schizophrenia}.
At the same time, tremendous technological advancements in serial-section electron microscopy are making the systematic investigation of synaptic connectivity at the level of detail of cortical microcircuits accessible~\cite{Bock:2011fk}.

Despite continuous progresses, the understanding of inter-relations between
the observed functional couplings and the underlying neuronal dynamics and circuit
structure is still a major open problem.
Several works have shown that functional connectivity \cite{Friston:2011iu} at
multiple scales is reminiscent of the underlying structural
architecture~\cite{Wang20131116,Honey2009Predicting, Deco2011Emerging}. This
structure-to-function correspondence is, however, not direct and is rather mediated by
interaction dynamics. On one side (``functional multiplicity''), structural networks generating a large
reservoir of possible dynamical states can give rise to flexible switching
between multiple functional connectivity networks~\cite{Battaglia:2012ff, Deco:2012jv}. On
the other (``structural degeneracy''), very different structural networks giving rise to analogous dynamical
regimes may generate qualitatively similar functional networks
\cite{Stetter:2012fe}. Therefore, particular care is required when interpreting
data originating from non-invasive functional data-gathering approaches such as fMRI~\cite{logothetiswhat2008}.  Altogether, these arguments call for highly controllable
experimental frameworks in which the results and predictions of different
functional connectivity analysis techniques can be reliably tested in different
dynamic regimes.

A first step in this endeavor consists in simplifying the neuronal system under
investigation. For this reason, different studies have focused on \emph{in
vitro} neuronal cultures of dissociated
neurons~\cite{Eckmann:2007ft,Wheeler2010Designing}.  Neuronal cultures are
highly versatile and easily accessible in the laboratory. Unlike in naturally
formed neuronal tissues, the structural connectivity in cultures can be dictated
to some extent~\cite{Wheeler2010Designing}, and even neuronal dynamical
processes can be regulated using pharmacological agents or optical or
electrical stimulation. These features have made neuronal cultures particularly
attractive for unveiling the processes shaping spontaneous activity,  including its initiation~\cite{Maeda:1995uu,Orlandi2013Noise}, synchronization~\cite{Eytan:2006hu} and plasticity~\cite{Wagenaar:2006fo,Cohen200821}, as well as 
self-organization~\cite{Pasquale20081354} and criticality~\cite{Tetzlaff:2010he}. Moreover, some studies also showed that spontaneous
activity in \emph{in vitro} preparations shares several dynamical traits with
the native, naturally formed neuronal tissues~\cite{Mazzoni:2007hb}.

A second step consists in developing and testing the analysis tools that
identify directed functional interactions between the elements in the network.
Information theoretic measures such as Transfer Entropy
(TE)~\cite{Kaiser:2002gc,Schreiber:2000jx} can capture linear and non-linear
interactions between any pair of neurons in the network. TE does not require
any specific interaction model between the elements, and therefore it is
attracting a growing interest as a tool for investigating functional
connectivity in imaging or electrophysiological studies~\cite{Gourevitch:2007gp,
Besserve:2010ge, Wibral2011TE-MEG,Vicente:2011ix}. The independence of TE on
assumptions about interaction models has made it adequate to deal with different
neuronal data, typically spike trains from simulated
networks~\cite{Kobayashi:2013gw}, multi-electrode
recordings~\cite{Bettencourt2007Functional,Garofalo2009Evaluation,Ito:2011wc,Marconi2012Emergent}
or calcium imaging fluorescence data~\cite{Stetter:2012fe}. TE proved to be successful
in describing topological features of functional cortical cultures~\cite{Bettencourt2007Functional,Garofalo2009Evaluation,Marconi2012Emergent},
and in reconstructing structural network connectivity from activity~\cite{Ito:2011wc,Stetter:2012fe}.

In a previous work~\cite{Stetter:2012fe}, we investigated the assessment of
excitatory-only structural connectivity from neuronal activity data (with
inhibitory synaptic transmission blocked). For this purpose we developed an extension
of TE, termed Generalized Transfer Entropy (GTE). To test the accuracy of our
connectivity reconstruction method, we considered realistic computational models
that mimicked the characteristically bursting dynamics of spontaneously active
neuronal cultures. Comparing diverse reconstruction approaches, we concluded
that GTE performed superiorly, even when systematic artifacts such as light
scattering were explicitly added to our surrogate data. Besides the inclusion of
corrections coping with the poor temporal resolution of typical calcium fluorescence
recordings, a key ingredient making GTE successful was dynamical
state selection, i.e. the restriction of the analysis to a dynamical regime in
which functional interactions were largely determined by the underlying hidden
structural connectivity. In particular we showed that it was necessary to
restrict the analysis to inter-burst regimes, while consideration of bursting epochs led to inference of exceedingly clustered structural
topologies~\cite{Stetter:2012fe}.

Here we extend our previous work, by attempting the inference of both excitatory and
inhibitory connectivity. Inhibition is a
major player in regulating neuronal network dynamics, and the regulation of the
excitatory-inhibitory balance is crucial  for optimal circuit function~\cite{Poil:2012er,Lombardi:2012ia}. In the brain, inhibition shapes
cortical activity~\cite{Isaacson2011Inhibition}, dominates sensory responses
\cite{Haider2012Inhibition}, and regulates motor behavior~\cite{Arber2012Motor}.
Severe behavioral deficits in psychiatric diseases such as autism and
schizophrenia have been ascribed to an imbalance of the excitatory and
inhibitory circuitry~\cite{Yizhar2011Neocortical_E-I}. Despite the importance of
inhibition, functional connectivity studies often disregard it because of the
difficulty in its identification. Hence, unraveling inhibitory connections, and
their interplay with the excitatory ones in shaping network dynamics, is of
major interest. We show here that the TE-based approach that we previously used for the
inference of excitatory connectivity can be extended with virtually no
modifications to networks including as well inhibitory interaction, whose
dynamics is once again reproduced by realistic computational models for which
the ground-truth connectivity is known. We reveal that the most
difficult inference problem is not the identification of a link, be it
excitatory or inhibitory, but rather the correct labeling of its type.
We show that an elevated accuracy of labeling of both excitatory
and inhibitory links can be obtained by combining the analysis of network
activity in two conditions, a first one where both excitation and inhibition are
active, and a second one where inhibition is pharmacologically removed. We show as well, however, 
that the inference of link types remain extremely uncertain with current experimental protocols. As a perspective solution, we foresee, based on extensive simulations, that significant
improvements in both reconstruction and labeling performance could be
achieved by enhancing the spontaneous firing of a culture through a weak
external stimulation. 

\section*{Results}

\subsection*{Dynamics of biological and simulated networks}

\begin{figure}[!ht] \begin{center}
    \includegraphics[width=17.35cm]{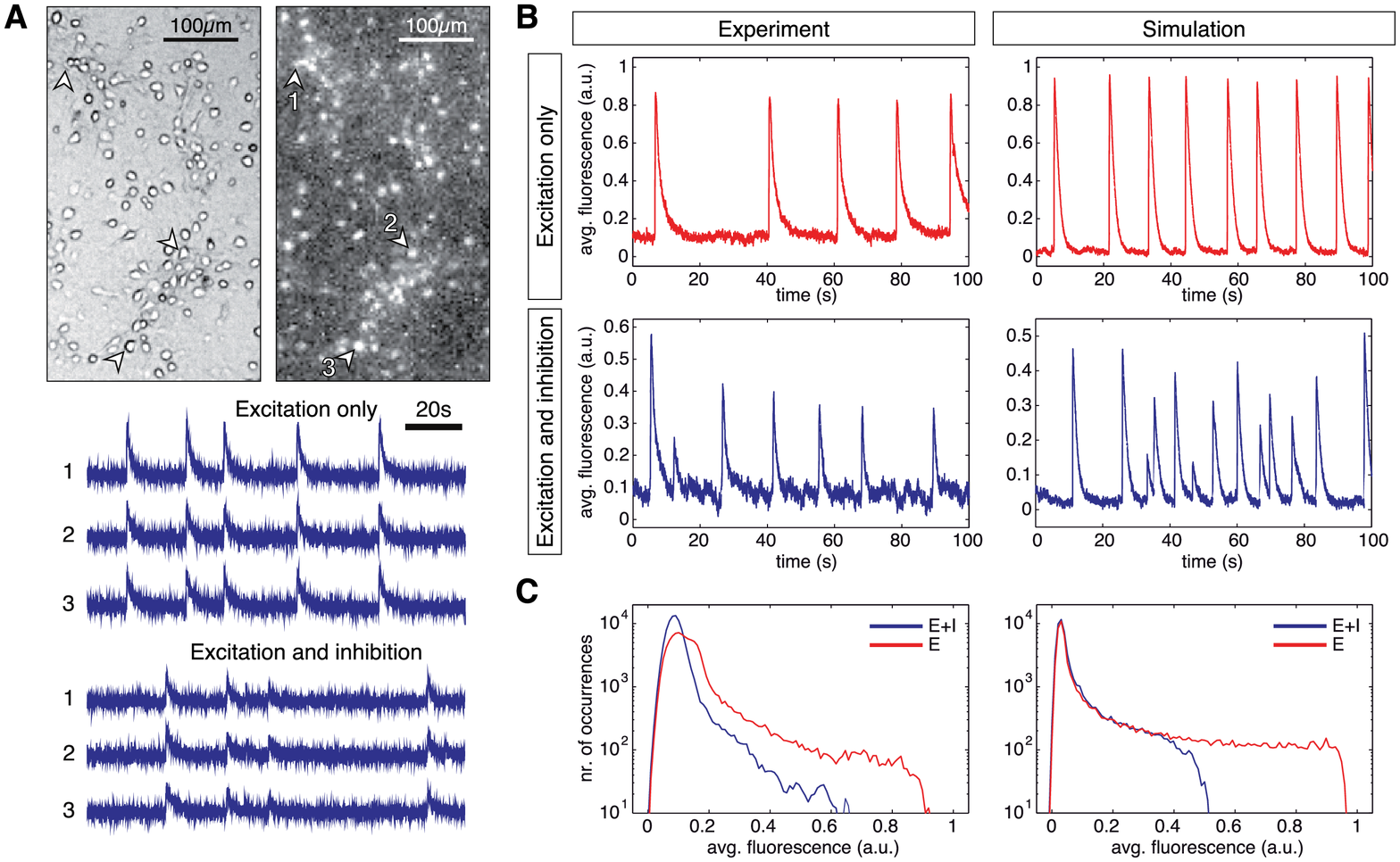}
  \end{center}
  \caption{\textbf{Neuronal network dynamics.}
    \textbf{A} Top: Bright field and fluorescence images of a small region of a neuronal culture at day \emph{in vitro} 12. Bright spots correspond to firing neurons. Bottom: Representative time traces of recorded fluorescence signals of 3 individual neurons. The numbers beside each trace identify the neurons on the images. Data shows, for the same neurons, the signal in recordings with only excitation active (``E'') and the signal with both excitation and inhibition active  (``E+I'').
    \textbf{B} Population-averaged fluorescence signals in experiments (left)
    and simulations (right), illustrating the semi-quantitative matching between \emph{in vitro} and \emph{in silico} data. Top: excitatory-only traces (``E--only'' data). For the experiments, inhibition was silenced through application of saturating concentrations of bicuculline. For the simulations, inhibitory synapses were silenced by setting their efficacy to zero. Bottom: traces for both excitation and inhibition active (``E+I'' data).  Network bursts appear as a fast increase of the fluorescence signal followed by a slow decay. Bursts are more frequent and display lower and more heterogeneous amplitudes in the presence of inhibitory connections.
    \textbf{C} Histogram of population-averaged fluorescence intensity for a 1
    h recordings in experiments (left) and simulations (right). Data is shown in semilogarithmic scale for clarity. Red curves correspond to the ``E--only'' condition, and the blue curves to the ``E+I'' one.
  \label{fig:1}}
\end{figure}

Dissociated neurons grown \emph{in vitro} self-organize and connect to one
another, giving rise to a spontaneously active neuronal network within a week
(see Figure~\ref{fig:1}A)~\cite{Eckmann:2007ft,Soriano:2008ic,Chiappalone200641,Cohen200821}.
About 70--80\% of the grown connections are excitatory, while the remaining
20--30\% are inhibitory \cite{Soriano:2008ic}. Activity in neuronal cultures is
characterized by a bursting dynamics, where the whole network is active and
displays quasi-synchronous, high frequency firing within 100--200 ms windows
\cite{Cohen200821}. The timing of the bursts themselves is irregular, with
average inter-burst intervals on the order of 10 s in a typical preparation. Between different bursts, firing across the network has a low-frequency and can be described as asynchronous.

Neuronal dynamics in cultures may be monitored using calcium fluorescence
imaging (see Methods)\cite{Eckmann:2007ft,Grienberger2012862}, which enables
the recording of the activity of thousands of individual neurons
simultaneously.  Figure~\ref{fig:1}A shows example traces illustrating the
characteristic fluorescence signal of individual neurons \emph{in vitro}. The fluorescence
signal is characterized by a fast onset as a result of neuronal activation and the
binding of $\text{Ca}^{2+}$ ions to the fluorescence probe, followed by a slow decay back to
the baseline due to the slow unbinding rate.  This behavior is apparent in the
population average of the signal, as shown in Fig.~\ref{fig:1}B, where bursts
are clearly identified by the fast rise of the fluorescence signal.

To appraise the role of inhibition on dynamics, we monitor neuronal network activity in two different conditions:
A first one, with only excitatory connections active, where inhibitory connections
have been completely blocked (denoted as ``E--only'' networks); and a second
one, where both excitatory and inhibitory connections are functionally active (herein after
denoted as ``E+I'' networks).  In experiments, inhibitory synapses are
silenced through the application of saturating levels of bicuculline, a
GABA$_{\text{A}}$ receptor antagonist (see Methods). An example trace of the
population average signal of such an excitatory-only system is shown in the top
left panel of Fig.~\ref{fig:1}B, whereas the dynamic behavior in presence of inhibition is shown in
the bottom left panel of Fig.~\ref{fig:1}B. In the ``E--only'' condition,
bursts are more pronounced and more regular in amplitude than in the ``E+I'' condition, an effect also seen
in other studies~\cite{Cohen200821,Jacobi2009,Tibau:2013hz}.

These recordings in neuronal cultures provide a comparison reference for our simulated networks of model neurons.
We build a computational model of a culture whose dynamics capture its major
qualitative features. These include a high variability in the inter-burst intervals, a low $\sim0.1$ Hz inter-burst firing rate, and, in presence of inhibition, an increase
in bursting frequency as well as a general decay in the amplitudes of the fluorescence signal, paired by an increase in their heterogeneity. More specifically, we consider a network of $N = 100$ leaky integrate-and-fire nodes with
depressive synapses in combination with a model for the calcium fluorescence. Network connectivity is random and sparse, with links rewired in order to reach an above-chance level of clustering (see Methods).
Each node receives inputs from its pre-synaptic neighbors as well as from
independent external sources to mimic spontaneous single neuron activity due to noise
fluctuations in the ionic current through its membrane. Free model parameters, such as the homogeneous conductance weights of recurrent connections, were calibrated
such as to yield dynamics comparable to the biological recordings, with a bursting rate of 0.1Hz and realistic decay time constants of the calcium fluorescence (see the bottom right panels of
Figure~\ref{fig:1}B).  The blocking of inhibitory connections (top right panel
of Figure~\ref{fig:1}B) is simulated by setting the synaptic weight of all
inhibitory connections to zero (note, therefore, that the firing itself of inhibitory neurons is not suppressed, but just its postsynaptic effects).

As discussed more in depth in~\cite{Stetter:2012fe}, a  hallmark of bursting dynamics is the right-skewed histogram of the population average of the calcium fluorescence signal (see Figure~\ref{fig:1}C). Low
fluorescence amplitudes are associated to the non-bursting regime, which is noise
dominated, and the right tail of the distribution reflects bursting events.
The range spanned by this right tail is distinctly shortened in presence of inhibition.
This difference in the large fluorescence amplitude distribution can be ascribed to the dynamics at the synapse level: For purely excitatory networks, the neurotransmitters resources of a given synapse
are depleted during a bursting event~\cite{Cohen:2011bi}. Neurons experience
high frequency discharge, but require a longer time
to recover, giving rise to long inter-burst intervals. Inhibition lowers this release of neurotransmitters by
suppressing neuronal firing before complete depletion, therefore providing a
faster recovery, shorter inter-burst periods and lower firing activity inside
the bursts.

\subsection*{Reconstructing structural connectivity from directed functional links}

Based on simulations of the calcium dynamics in the network, a network of (directed) functional connectivity is reconstructed by computing the Generalized
Transfer Entropy~(GTE) for each (directed) pair of links (see Methods).
GTE is an extension of Transfer Entropy, a measure that quantifies predictive information flow between stationary systems evolving in time~\cite{Schreiber:2000jx}. As an information theoretical implementation of the Granger Causality concept \cite{Granger:1969wx}, a positive TE score assigned to a directed link from a neuron $i$ to a neuron $j$ indicates that the future fluorescence of $j$ can be better predicted when considering as well the past fluorescence of $i$ in addition to the past of $j$ itself. We previously introduced GTE to study
the reconstructed topology of purely excitatory networks under diverse network dynamical
states and signal artifacts~\cite{Stetter:2012fe}. Here we extend its applicability to data that includes inhibitory action.

\subsubsection*{Conditioning as state selection}

A central observation that motivated the definition of GTE was the existence of
different dynamical states in the switching behavior from asynchronous firing
to synchronous bursting activity. The distribution of fluorescence amplitudes
(see Figure~\ref{fig:1}C) provides a visual guide to the relative weight of the
single activity events and the bursting episodes. A functional reconstruction
in this bursting regime shows a very clustered connectivity due to the tightly synchronized firing of large communities of neurons. We can understand intuitively this finding, by considering that, in the bursting regime, the network is over-excitable and the firing of a single neuron can trigger the firing of a large number of other neurons not necessarily linked to it by a direct synaptic link. On the other hand, the neuronal activity in the
non-bursting regime is sparse and dominated by pairwise interactions, and thus,
a reconstruction in this regime identifies directed functional interactions that more closely
match the structural connectivity (i.e. high GTE might signal direct pre- to post-synaptic coupling in this regime), as previously discussed thoroughly for ``E--only'' networks~\cite{Stetter:2012fe}.

\begin{figure}[!ht]
  \begin{center}
    \includegraphics[width=12.35cm]{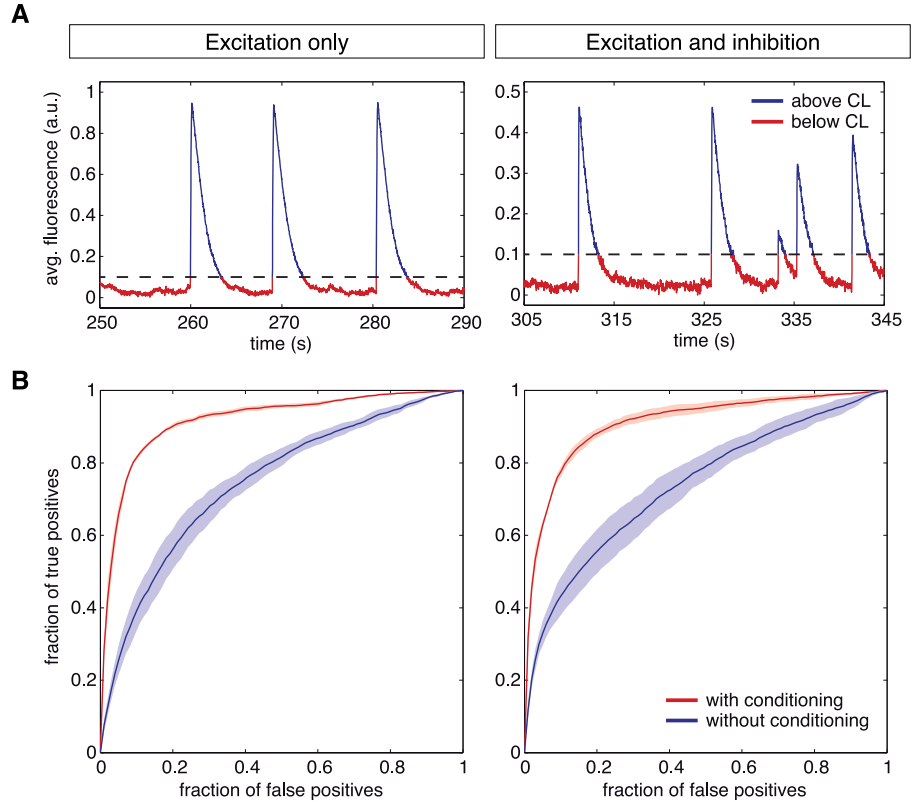}
  \end{center}
  \caption{\textbf{Signal conditioning.}
    \textbf{A} Separation of the signal in two regimes according to the conditioning level (dotted line), a first one that encompasses the low activity events (red curves), and a second one that includes the bursting regimes only (blue). The same conditioning procedure is applied in both ``E--only'' networks (left) and in ``E+I'' ones (right).
    \textbf{B} \emph{Receiver Operating Characteristic} (ROC) curves quantify the accuracy of reconstruction and its sensitivity on conditioning. Functional networks are generated by including links with a calculated GTE score exceeding an arbitrary threshold. ROC curves plot then the fraction of true and false positives in the functional networks inferred for every possible threshold. For ``E--only'' networks (left) and ``E+I'' networks (right), the red curves show the goodness of the reconstruction after applying the conditioning procedure. Blue curves illustrate the reconstruction performance without conditioning. The ROC curves show that the conditioning procedure significantly improves reconstruction performance. ROC curves were averaged over different network realizations (95\% confidence intervals shown).
    \label{fig:2}}
\end{figure}

A rough segmentation of the population signal into time sequences of bursting and
non-bursting events is simply achieved by defining a fixed \emph{conditioning level}
on the population average fluorescence. This simple modification with respect to the original TE formulation, makes GTE suitable for an analysis of functional interactions which distinguish different dynamical regimes, as illustrated for purely
excitatory networks in the left panel of Figure~\ref{fig:2}A. The network is indeed considered to be in a bursting regime when the network-averaged fluorescence exceeds the chosen conditioning level (dotted line in  Figure~\ref{fig:2}A), and in an inter-burst regime otherwise. The value of the
conditioning level itself is obtained through the analysis of the fluorescence
signal histogram and set close to the transition from the Gaussian-like
profile shown for low fluorescence values to the long tail characteristic of
the population bursts.

Note that, while our approach works by restricting the analysis to epochs of inter-burst activity only, other complementary methods exploit detailed information about typical burst build-up sequences in order to infer structure, with potentially superior results when the required time resolution is accessible (e.g. \cite{Pajevic:2009hj}).

\subsubsection*{Connectivity reconstruction of simulated ``E--only'' networks}

Reconstruction performances from the GTE computation are quantified in the form
of \emph{receiver operating characteristic}~(ROC) curves. These curves are
obtained as follows: GTE assigns a score to every possible link in the network,
and only scores above a given threshold are considered as putative links. These
accepted links are then systematically compared with the ground truth topology
of the network, and for gradually lower threshold levels. The ROC curves
finally plot the fraction of true positives, i.e., inferred connections which really exist, as a
function of the fraction of false positives, i.e., wrongly inferred connections.

The ROC curves of the reconstruction performance, with and without conditioning,
for the case of simulated ``E--only'' networks are shown in the left panel of
Figure~\ref{fig:2}B. Without conditioning (blue ROC curves), the reconstruction
quality of excitatory connections ---~to both excitatory and inhibitory neurons
confounded~--- is significantly better than a random choice (which would
correspond to a diagonal line in the ROC curve). The reconstruction is, however, hindered by the fact that the analysis effectively averages over data from
multiple dynamical regimes as described above. The
reconstruction performance thus significantly increases by applying a conditioning (red ROC curves) which selects uniquely the inter-burst regime.

It was also shown for simulations comparable to the ones generated as described
above, that the reconstructed networks using GTE are approximately unbiased
regarding bulk network properties, such as the mean clustering coefficient, or
the average length of connections in the network~\cite{Stetter:2012fe}.

\subsubsection*{Connectivity reconstruction of simulated ``E+I'' networks}

An important aspect of Transfer Entropy, and by extension of GTE, is its
model-free nature. Thus, during the process of identifying causal influences
between neurons, there is no need to define a generative model for neuronal
firing or calcium dynamics, as in the case, e.g., of Bayesian inference approaches \cite{Mishchenko:2011ui}. It follows that we can apply GTE without modifications to the case in which both excitatory and inhibitory links are active, provided that the inter-burst network state can be identified in an analogous way.  Indeed we
observe that while the presence of inhibition does change the dynamics of the
system to some extent, the switching behavior remains robustly present
(see the right panel of Figure~\ref{fig:2}A), allowing the straightforward identification of a performing conditioning level.

Remarkably, the reconstruction performance of ``E+I'' networks remains at high
levels after conditioning, of about 80\% true positives at 10\% false
positives, as shown in the right panel of Figure~\ref{fig:2}B. Thus the
model-independence of GTE allows the reconstruction of both excitatory and
inhibitory links. As a further self-consistency check, we have simulated the
dynamics of a neuronal culture with a topology identical to the inferred one and
compared it with the dynamics of the network with the original ground-truth
topology. The resulting bursting and firing rates, for both the ``E-only'' and
the ``E+I'' cases, are not statistically significantly different from the case
of perfect reconstruction, while they markedly differ from the case of a randomized topology (not shown). Nevertheless, given the phenomenon of structural degeneracy, a large number of even very different structural circuits could give rise to equivalent dynamical regimes \cite{Stetter:2012fe}. Therefore, passing this self-consistency check is not a sufficient condition to prove high reconstruction quality, though it is a necessary one.

Note, finally, that we have disregarded, until now, the identification of the specific type, i.e. excitatory or inhibitory, of each link, focusing uniquely on whether a link is present or absent in the ground-truth structural network, whatever is its nature. As previously mentioned, correctly labeling a link turns out to be a more elaborated task than just inferring its existence.

\subsection*{Distinguishing excitatory and inhibitory links}

GTE probes the existence of unspecified influences between signals, but cannot
identify the type of occurring interaction \emph{a priori}. Its versatility
also means that very different types of interactions can give the same GTE
score if their influence in terms of predictability is the same.  Hence, to
separate between excitatory and inhibitory connections we have to either
introduce \emph{ad hoc} information on neuronal types or combine different
reconstructions together to single out the different connectivity types.

Such \emph{ad hoc} information might come from
dye impregnation, fluorescence labeling or
immunostaining~\cite{lichtmana2008Technicolour}. These techniques identify cell
bodies and processes according to some specific traits, for instance membrane
proteins or neurotransmitters' receptors. According to Dale's
principle~\cite{kandel1967Dale}, a neuron shows the same distribution of
neurotransmitters along its presynaptic terminals. Hence, if a neuron is labeled as either excitatory or inhibitory, we
can assume that all its output connections are of the same matching type.
Thus by combining the type of information provided by some extrinsic labeling
technique with the unspecific causal information provided by GTE, the
overall set of inferred links can be separated into two non--overlapping subsets
of excitatory and inhibitory links.

Being able to identify the type of a neuron ---~even with perfect
accuracy~--- does not guarantee \emph{a priori} that excitatory and inhibitory
links can be inferred equally well. On the contrary, different reconstruction
performances have to be expected in general, since the interaction mechanism of
excitatory links is inherently different from the inhibitory ones, the former
promoting the activity of the target neuron, whereas the latter restrain it.
We have tested the accuracy of this \emph{ad hoc} approach through numerical
simulations. GTE is applied to the ``E+I'' data, and the reconstruction quality
is assessed separately for the connections originating from neurons of different
types (see Methods). Non trivially, the results of this analysis indicate that
both types of connections are reconstructed with high accuracy (see
Figure~\ref{fig:3}A). At a fraction of 10\% of false positives, excitatory links
are detected at a true positive rate of 80\%. Inhibitory links show a lesser but
still high detection accuracy, of about 60\% of true
positives.

\subsection*{Reconstructing and labeling connections from spontaneous dynamics}

In the absence of information on neuronal types, an alternative
approach consists in a direct combination of the reconstructions procured by
the ``E--only'' and ``E+I'' data on the same neurons. By adding together the GTE scores from the two
reconstructions we can assume that the higher scores come from links that show a
high score in both reconstructions. This procedure is thus expected to highlight the pool of excitatory
connections, since they are the only ones present in both network
conditions. Similarly, we can subtract the ``E--only'' scores
from the ``E+I'' ones. High scores will then now highlight those links that are present in the ``E+I'' but not in the ``E--only'' network, i.e. the pool of inhibitory connections.

\begin{figure}[!ht]
  \begin{center}
    \includegraphics[width=12.35cm]{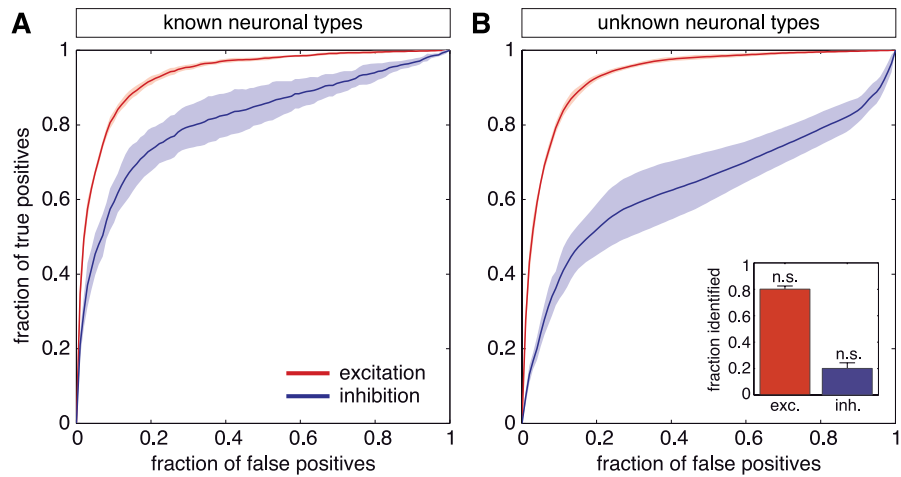}
  \end{center}
  \caption{\textbf{Optimal network reconstruction.}
    \textbf{A} ROC curves for the reconstruction of a network with both
    excitatory and inhibitory connections active, supposing to know \emph{a priori}
    information about neuronal type. GTE is first applied to the ``E+I'' data. Next,
    following Dale's principle and exploiting the available information on neuronal type, links are classified according to their excitatory (red) or inhibitory (blue) nature.
    \textbf{B} ROC curves for the best possible identification of excitatory
    and inhibitory connections, when information on
    neuronal type is unaccessible. Excitatory links (red) are identified by adding together
    the Transfer Entropy scores of simulations run in ``E--only'' and ``E+I'' conditions, and later thresholding them. Inhibitory
    links (blue) are identified by computing the difference in Transfer Entropy
    scores between the runs with inhibition present and blocked.
    Inset: fraction of excitatory and inhibitory neurons correctly identified
    from these ROC curves. Results were not significantly different from random guess (see Methods).
    All the results were averaged over  different network realizations. The shaded areas in the main plots, as well as the error bars in the inset, correspond to 95\% confidence intervals.
  \label{fig:3}}
\end{figure}

The performance of this first two-step reconstruction approach is shown in
Figure~\ref{fig:3}B. The reconstruction of excitatory connections has a quality as good as the one obtained with \emph{a priori} knowledge
of neuronal type based on extrinsic labeling (see Figure~\ref{fig:3}A). However, the performance markedly deteriorates for the reconstruction of inhibitory links, since only 40\% of the inhibitory connections are correctly identified at 10\% of false positives.

Note that an additional complication arises with the described two-steps
pipeline. A given link might be attributed a
combined score above the inclusion threshold, both when considering the sum
\emph{and} the difference of original GTE scores. In this case, the link would
be labeled as ``both excitatory and inhibitory'', a fact which is excluded by
Dale's principle. Despite this problem, we might still try to combine the ``E--only'' and ``E+I'' reconstructions to infer the nature of each neuron. To test the accuracy of such identification we try to label neurons as excitatory or inhibitory based on a highly ``pure'' structural network reconstruction. To do so, we select a very high GTE threshold for link, in such a way that in the inferred subnetwork ---including, correspondingly, very few links only--- the fraction of false positives remains small (with a maximally tolerable ratio of 5\%). We first sum and subtract ``E--only'' and ``E+I'' scores to obtain putative
excitatory and inhibitory links, as just discussed. We next compute the output
degrees of the neurons for each subnetwork, $k_E$ and $k_I$, respectively.
Finally, we rank each neuron according to the difference~$k_E-k_I$. Following
Dale's principle, the set of neurons with the highest (positive) ranking would be labeled as excitatory, and those with the lowest (negative) ranking as inhibitory. The results, however, as shown in the inset
of Figure~\ref{fig:3}B, indicate that this approach does not provide better results than
a random guessing of neuronal type (see Methods for details on significance
testing) and a different approach is required.

\subsection*{Reconstructing and labeling connections from stimulated dynamics}

As a matter of fact, the major challenge for an accurate reconstruction and precise labeling of neuronal types is the identification of inhibitory links, and this for the following reason.
To estimate GTE, we need to evaluate the probability of each given neuron to be active in a short time window of a duration $\Delta t = (k+1) \; \tau_{\text{image}}$, where $k=2$ is the order of an assumed Markov approximation (see Methods) and $\tau_{\text{image}} = 20 \text{ms}$ is the image acquisition interval. With these parameter choices, we obtain then $\Delta t = 60 \text{ms}$. Neurons in a culture spike with an average inter-burst frequency of $\nu \sim 0.1\text{Hz}$, resulting in a low firing probability within each time bin.
Continuing this reasoning,  the probability that two unconnected neurons spike at random in the same time window is given by $(\nu \, \Delta t)^2 \sim 4\cdot10^{-5}$.
The number of coinciding events~$N_{\text{events}}$ expected in a recording is thus:
\begin{equation}
N_{\text{events}} \sim N_{\text{samples}} \, (\nu \, \Delta t)^2 ,
  \label{eq:nevents}
\end{equation}
where $N_{\text{samples}}$ is the number of independent samples in a
recording.
In  a typical recording session lasting~$\sim 1$h, one gets $N_{\text{samples}} \sim 1.8 \cdot 10^{5}$ independent samples and therefore $N_{\text{events}} \sim 6$.  Hence, one can expect to observe, on average, just six concurrent spikes between any pair of unconnected neurons.
If an excitatory link exists between two neurons, the conditional
probability of firing rises above this random level and more coincidence events
are observed, turning into an appreciable contribution to the GTE calculation. However, if an inhibitory link is present, the number of simultaneous spikes gets further reduced with respect to the already very small chance level, making any accurate statistical assessment very difficult. Nevertheless, we note that the number of detected events scales as $\nu^2$ with the frequency
of firing, and even a slight increase in spiking frequency would enhance
considerably the reconstruction performance.

A promising approach to increase neuronal firing consists in forcing the
neuronal network through external stimulation. Several studies on neuronal cultures have used
external drives, typically in the form of electrical stimulation, to act on
neuronal network activity, for instance to investigate connectivity
traits~\cite{Breskin2006Percolation,Soriano:2008ic}, modify or control activity
patterns~\cite{Wagenaar2005Controlling,vanPelt2005Dynamics}, or explore network
plasticity~\cite{Madhavan2006Multisite,Wagenaar2006Searching}. Such \emph{in vitro}
approaches are reminiscent of \emph{in vivo} clinically relevant techniques such as
deep brain stimulation, used in the treatment of epilepsy and movement
disorders~\cite{McIntyre2004Uncovering,Durand2001Suppression}.

External stimulation in neuronal cultures has been reported to increase
neuronal firing~\cite{vanPelt2005Dynamics} and to reduce network
bursting~\cite{Wagenaar2005Controlling,Madhavan2006Multisite}, a combination of
factors that, in the GTE reconstruction context, improve the accuracy in the
identification of the network architecture. To explore potential improvements in
reconstruction, we simulate the effect of an applied external drive in a purely
phenomenological way by increasing the frequency parameter of the Poisson
process that drives spontaneous activity. This additional drive never increases
the spontaneous firing frequency beyond 3 Hz, being meant to represent the effects
of a rather weak external stimulation. Due to this contained increase of firing
rate, the collective bursting activity of the simulated network continues to be shaped dominantly by network interactions, rather than by the drive itself.

\begin{figure}[!ht]
  \begin{center}
    \includegraphics[width=12.35cm]{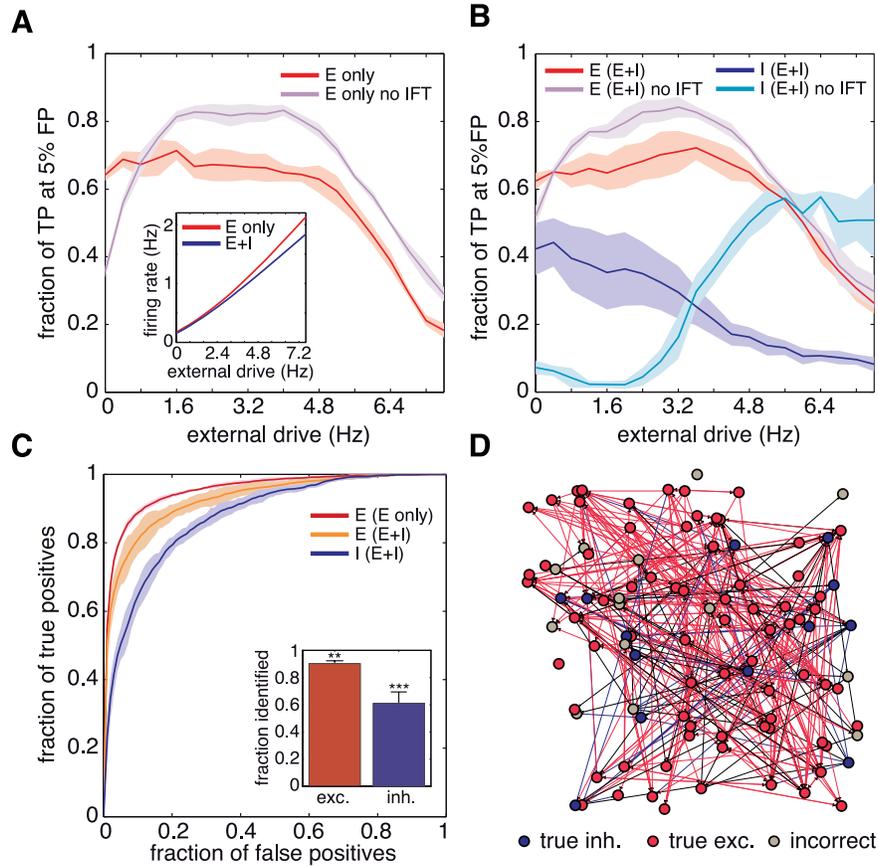}
  \end{center}
  \caption{\textbf{Reconstruction improvement through external stimulation.}
    \textbf{A} and \textbf{B}, fraction of true positives from the
    reconstructions at the 5\% false positive mark for the studied networks.
    ``E--only'' networks are shown in \textbf{A}; ``E+I'' networks in \textbf{B}. Inset: dependence of the
    spontaneous firing rate on the applied external drive, emulated here by increasing the rate of the background drive to the culture \emph{in silico}. All the excitatory
    reconstructions reach a stable plateau in the reconstruction after removal of the
    instantaneous feedback term~(IFT) correction (see Methods). The inhibitory reconstruction is accurate only
     for higher values of the external drive.
    \textbf{C} ROC curves extracted from \textbf{A} and \textbf{B}
    with an external stimulation of 4 Hz. Inset: fraction of excitatory and inhibitory
    neurons correctly identified from these reconstructions. Identification was statistically significant compared to
    random guessing. For excitatory neurons,  $p<0.01$ (**); for inhibitory neurons, $p<10^{-4}$  (***).
    \textbf{D} Example of an actual reconstruction after identification of neuronal type. Identified excitatory neurons
    are shown in red and inhibitory ones in blue. Incorrectly identified neurons are shown in grey.
    Correctly identified excitatory and inhibitory links are shown in red and
    blue, respectively, and wrongly identified links are shown in black. For clarity in the representation of the links, a threshold value lower than the optimal has been applied.
  \label{fig:4}}
\end{figure}

The performance of our GTE algorithm combined with a weak network stimulation is illustrated in Fig.~\ref{fig:4}A, where we show the fraction of true positives in the reconstruction of ``E--only'' networks at 5\%
false positives. The presence of even very small external drives substantially
enhances reconstruction based on GTE. For higher drives, reconstruction performance reaches a plateau that quantifies the range of optimum stimulation. Performance later decays due to the
excess of stimulation, which substantially perturbs spontaneous
activity and alters qualitatively the global network dynamics.
We incidentally remark
that the incorporation of the external drive makes unnecessary ---~actually, even
deleterious~--- the instantaneous feedback term correction (IFT, see Methods),
i.e., an \emph{ad hoc} modification to the original formulation of TE which was
introduced in~\cite{Stetter:2012fe} to cope with the poor frame rate of calcium
fluorescence recordings, definitely slower than the time-scale of monosynaptic
interaction delays. The IFT correction allows to encompass interactions occurring in the same temporal bin of the recording for TE estimation, a feature useful to enhance reconstruction results when the time-scale of pre-postsynaptic neuron interactions is fast relative to the time resolution of the recording. However, same--bin interactions also result in an overestimation of bidirectional connections, since one cannot establish directionality within a single time bin. When the firing rate is enhanced with respect to spontaneous conditions these negative effects of the IFT corrections become predominant.

The same reconstruction analysis for ``E+I'' networks is shown in
Fig.~\ref{fig:4}B, for excitatory and inhibitory links separately. The
identification of excitatory links greatly improves with moderate drives and,
again, IFT becomes unnecessary. For inhibitory links, performance is
optimum at low drives, when IFT is used. Without IFT, however, performance is better at relatively
high drives, and one can observe the existence of an optimal stimulation range
(leading to a firing rate of~$\sim5$Hz) that maximizes inhibition
reconstruction while preserving a relatively good excitatory identification. 

We note as well that, for ``E+I'' networks, bursts disappear in general at
higher values of the external drive.  In general, as depicted in the inset of
Fig.~\ref{fig:4}A, the dependence of the spontaneous firing frequency on the
external drive is quantitatively different from ``E--only'' networks, requiring
typically a stronger drive to achieve a comparable firing rate.

With the external drive the overall ROC curves are also improved. In
Figure~\ref{fig:4}C we show the reconstruction performance for medium values of
stimulation. In this new regime, we can again try to determine the neuronal type
based on the labeling procedure used in the previous section (inset of
Figure~\ref{fig:4}C). Now excitatory neurons are correctly identified with 90\%
accuracy, whereas the fraction of inhibitory neurons correctly identified rises
conspicuously to 60\%. This marked improvement is now statistically significant
(see Methods). 

In Figure~\ref{fig:4}D we show an actual reconstruction of a portion of the
original network with this procedure. Correctly inferred excitatory and
inhibitory neurons are shown in red and blue respectively, and mismatches in
yellow. Correctly identified excitatory and inhibitory links are also shown in
red and blue respectively, and false positives are shown in black. It is
visually evident that for this thresholding level a very high purity is
achieved, and only a small fraction of the reconstructed links are false
positives.

We conclude that the addition of a weak external stimulation to the ``bare''
network dynamics results in an overall improvement on the reconstruction of
both excitatory and inhibitory links. Moreover, by combining the reconstructions of
``E--only'' and ``E+I'' networks, we also become able to infer the neuronal type by just
analyzing the dynamics, with no \emph{a priori} knowledge of the system and without resorting to extrinsic information of any sort.

\section*{Discussion}

Living neuronal networks contain both excitatory and inhibitory neurons.
Although the interplay between excitation and inhibition gives rise to the rich
dynamical traits and operational modes of brain circuits, inhibition is often
neglected when analyzing functional characteristics of neuronal circuits, mostly
because of its difficult identification and treatment. In this work we have made
a first step towards filling this gap, and introduced a new algorithmic approach
to infer inhibitory synaptic interactions from multivariate activity
time-series. In the framework of a realistically simulated neuronal network that
mimics in a semi-quantitative way key features of the behavior of
neuronal cultures, we applied Generalized Transfer Entropy (and Dale's principle) to identify excitatory as well as inhibitory connections and neurons.

In a previous work~\cite{Stetter:2012fe}, we developed the GTE framework and
applied it to extract topological information from the dynamics of purely
excitatory networks, but left as an open question the treatment of inhibition.
Here we have shown that GTE has the potential to be applied without substantial modifications to
recordings relative to cultures with active inhibition (``E+I'' cultures). This
data is characterized by an irregular bursting dynamics with overall lower
---~but distinctly fluctuating~--- fluorescence amplitudes as well as higher
bursting frequencies than purely excitatory (``E--only'') signals. In general,
GTE provided an overall good reconstruction of the ``E+I'' simulated data, hinting at the
robustness and general applicability of the algorithm. This is a highly non
trivial achievement of the algorithm, given the profoundly different functional
profile of inhibitory actions. The GTE reconstruction alone performed
well in identifying the existence of links between pairs of neurons, however, it was not
sufficient to resolve their excitatory or inhibitory nature. Yet, we provided
evidence through numerical experiments that this additional goal could be
fulfilled by retrieving \emph{a priori} information about the types of
different neurons (e.g. through immunostaining or selective fluorescent dyes),
or by combining the reconstructions obtained from both ``E+I'' and ``E--only''
recordings from a same network (thus, again relying uniquely on time-series
analysis).

When \emph{a priori} information about the type of each neuron is available,
Dale's principle proves to be, at least in our simulations, a solid
yet simple approach that allows the identification of the major connectivity
traits of the neuronal network. However, when applying Dale's
principle to actual, living neuronal networks recordings (see
later), one has to consider its possible limitations, like the existence of
(rare) exceptions to it~\cite{Nicoll1998DaleLawRevision}.
We also remark
that, in a more realistic context, other types of \emph{a priori} information
beyond the nature of the neurons and their processes could be considered, like,
e.g. information about their spatial distribution. Although in this work we have
considered only purely random distance-independent topologies, neuronal cultures
grow on a bi-dimensional domain, and excitatory connections are typically of
shorter range than inhibitory ones. This kind of information could be integrated
in the analysis of network models that include metric properties and accounts
for spatial embedding (such as~\cite{Eckmann:2010ii,Orlandi2013Noise,Rudiger2014}), as well as different connectivity rules for the
generation of excitatory and inhibitory sub-networks.

A systematic extrinsic labeling of neuronal types might be difficult to achieve in large culture experiments. When \emph{a priori} information is unavailable, our results show that the
combination of the reconstructions for ``E--only'' and ``E+I'' spontaneous
activity data fails at identifying robustly the inhibitory interactions.
Nevertheless, we find that the reconstruction performance of excitatory links
remains almost unchanged when inhibition is present, despite the fact
that inhibition may substantially alter excitatory interactions, and in turn
network dynamics, for instance through feedback and feedforward inhibitory
loops. The observation that excitatory links are still correctly reconstructed
in ``E+I'' data shows the robustness of the algorithm to the presence of
different interactions in the system. We remark that the main factor
determining the poor identification of inhibitory links is the weak firing rate
during inter-burst epochs. Since, in a nearly asynchronous regime of inter-burst firing, the action of a direct inhibitory link manifests itself by reducing below the already small chance level the probability of firing coincidence between the two connected neurons, the recording of a larger amount of inhibitory firing would
be required to improve the reconstruction of inhibitory couplings.
Although the recording duration can be increased at will in numeric
simulations, this is not the case for real experimental recordings, to which our
algorithm aims at being applied.

In our simulations, we naturally achieved to increase single neuronal firing activity, and therefore reconstruction
statistics through a weak external stimulation of
the network, with neither a significant disturbance in neuronal network dynamics
nor the need for substantially longer recordings. In many previous works resorting to external drives to stimulate
network activity, both experimental and theoretical, the applied stimulation was
supra-threshold, i.e. the stimulation triggered directly neuronal firing
\cite{Breskin2006Percolation,Soriano:2008ic,Jacobi2009,Cohen:2010iq,
Linaro2011Inferring}. Our approach raises instead network excitability by a weak external drive that
effectively increases activity without modifying the network intrinsic behavior,
in the direction of other experimental studies that stimulated multiple sites of
a neuronal culture via a multi-electrodes array, to either increase network
firing, reduce the occurrence of bursting episodes, or investigate plasticity
\cite{vanPelt2005Dynamics,Wagenaar2006Searching}. Interestingly, these works observed that a weak stimulation along few hours did not induce plastic effects, i.e. did not change network behavior, thus making our reconstruction strategy of immediate applicability in experimental recordings.

In the present work we have exhibited experimental data only for qualitative
comparison with fluorescence traces obtained from the numerical model.
The experimental data could be analyzed in principle without need of any
modification to the GTE formulation, but we found our present knowledge of the
experimental recordings insufficient to get reliable reconstructions. In
particular, we are lacking good estimates of the neuronal firing rate during the inter-burst periods, as well as the amount of fluorescence change caused by an action potential. The former does not allow to determine whether we expect enough events to make the
reconstruction of inhibitory links feasible (see Eq.~(\ref{eq:nevents})), while
the latter prevents the application of an optimal data discretization strategy that would reduce the minimal recording length needed for accurate results. Our study intends therefore to foster the future application of the workaround
strategies here explored in experiments \emph{in silico}, i.e., most notably:
(i) a weak external stimulation to increase spontaneous activity; and (ii) the
extrinsic labeling of excitatory and inhibitory neuronal cell bodies after the
recording (to provide at least a partial source of \emph{a priori} information)
to be used in synergy with our algorithmic approach.

Finally, our reconstruction algorithm has the potential to be immediately applied to the
  analysis of fluorescence data in experimental recordings that are not affected by the aforementioned limitations. In particular, \emph{in vivo} recordings and brain slice measurements \cite{Mao2001Dynamics,Brustein2003,Dombeck200743}
  display a much richer activity at the individual neuron level than in the \emph{in
  vitro} counterparts.
Recent works have highlighted the ability of high speed multi-neuron calcium imaging
  to access neuronal circuits \emph{in vivo}
  \cite{Stosiek10062003,kerr2008imaging,Grewe2010high}. Our methodology can thus be
  directly applied to these data, particularly in those investigations that
  target the role of inhibition \cite{Bonifazi04122009,marissal2012pioneer}, although systematic verification of the inferred connectivity (in absence of a known ground-truth structure) remains currently out of reach and validation is only possible at the statistical level.

\section*{Methods}

All procedures were approved by the Ethical Committee for Animal Experimentation of the University of Barcelona, under order DMAH-5461.

\subsection*{Calcium traces from \textit{in vitro} cultures}

Experimental traces of fluorescence calcium signals were obtained from rat
cortical cultures at day \emph{in vitro} 12, following the procedures described
in our previous work~\cite{Stetter:2012fe} and in other studies
\cite{Segal:1992uk,Soriano:2008ic,Orlandi2013Noise}. Briefly, rat cortical neurons
from 18--19-day-old Sprague-Dawley embryos were dissected, dissociated and
cultured on glass coverslips previously coated with poly--l--lysine.  Cultures
were incubated at $37^{\circ}$C, 95\% humidity, and $5\%$ CO$_2$. Each culture
gave rise to a highly connected network within days that contained on the order
of 500 neurons/mm$^2$. Sustained spontaneous bursting activity appeared by day
\emph{in vitro} $6-7$. Prior to imaging, cultures were incubated for 40 min in
recording medium containing the cell--permeant calcium sensitive dye Fluo-4-AM.
The culture was washed with fresh medium after incubation and finally placed in
a recording chamber for observation. The recording chamber was mounted on a
Zeiss inverted microscope equipped with a Hamamatsu Orca Flash 2.8 CMOS camera.
Fluorescence images were acquired with a speed of 50 frames per second and a
spatial resolution of 3.4 $\mu$m/pixel.

In a typical measurement, we first recorded spontaneous activity as a long image
sequence 60 min long. Both excitatory and inhibitory synapses were active in
this first measurement (``E+I'' network). We next fully blocked inhibitory synapses
with 40 $\mu$M bicuculline, a GABA$_\text{A}$ antagonist, so that activity was
solely driven by excitatory neurons (``E--only'' network). Activity was next measured
again for another 60 min. At the end of the measurements, images were analyzed
to to retrieve the evolution of the fluorescence signal for each neuron as a
function of time.

Note once again that, in this study, experimental fluorescence traces were used only as a guiding reference for the design of synthetic data in ``E--only' and ``E+I'' conditions, and were not analyzed to provide network reconstructions, given the limitations of current experimental protocols, highlighted in the Results and Discussion section.

\subsection*{\emph{In silico} model}

\subsubsection*{Network generation}

We randomly distributed $N=100$ neurons over a square area of 1 mm$^{2}$.
Neurons were labeled as either excitatory with probability $p_E=0.8$ or
inhibitory with $p_I=0.2$. A directed connection (link) was created between any
pair of neurons with fixed probability $p=0.12$, giving rise to a directed
Erd\H{o}s-R\'enyi network\cite{Albert:2002it}. The resulting network is defined
by the adjacency matrix $A$, whose entries $a_{ji}=1$ denote a connection from
neuron $j$ to neuron $i$ ($j\to i$). The average full clustering
coefficient of the network~\cite{Fagiolo:2007gy} is given by
\begin{eqnarray}
  \text{CC}=\left\langle\frac{\left(A+A^\text{T}\right)^3_{ii}}{2T_i}\right\rangle_i,
  \label{eq:clustering}
\end{eqnarray}
where $A^\text{T}$ is the transpose of $A$ and $\langle\rangle_i$ denotes average
over index $i$. $T_i$ is defined as
\begin{eqnarray}
  T_i=d_i^t\left(d_i^t-1\right)-2d_i^\leftrightarrow,
  \label{eq:clusteringT}
\end{eqnarray}
where $d_i^t$ is the total degree of node $i$ (the sum of its in-- and
out--degree) and $d_i^\leftrightarrow$ is the number of bidirectional links of node
$i$. The clustering coefficient of the network after its construction was $\sim 0.12$, a value that was then raised up to a target one of 0.5 by following the Bansal \emph{et
al.} construction~\cite{Bansal:2009kl}, as follows. Two existing links $a_{ij}$ and
$a_{kl}$ were first chosen at random, with $i\neq j\neq k\neq l$. These links were then replaced by $a_{il}$
and $a_{kj}$.  This step was repeated until the desired
clustering coefficient was finally reached within a tolerance of 0.1\%. 

This above-chance clustering level was generated to account for experimental observations of clustered connections in neuronal local circuits \cite{Perin:2011kp}. We do not perform here a systematic study of the impact of CC on reconstruction performance, referring the reader to Ref.~\cite{Stetter:2012fe} for this issue, in which CC-independent performance is demonstrated.

\subsubsection*{Network dynamics}
Neurons in the simulated culture were modeled as integrate-and-fire units, of the form
\begin{eqnarray}
  \tau_m\frac{dV_i}{dt}=-(V_i-V_r)+\frac{1}{g_l}\left(I^A_i+I^G_i+\eta\right),
  \label{eq:iaf}
\end{eqnarray}
where $V_i$ is $i$-th neuron's membrane potential and $V_r=-70\text{mV}$ its resting
value, $\tau_m=20\text{ms}$ is the membrane time constant, $g_l=50\text{pS}$ is the leak
conductance, $I^A$ and $I^G$ the excitatory (AMPA) and inhibitory
(GABA$_\text{A}$) input currents respectively, and $\eta$ a noise term. When the
membrane potential reaches the threshold value $V_t=-50\text{mV}$ the neuron fires and
its membrane potential is reset to a value $V_r=-70\text{mV}$, which is maintained 
for a refractory time $\tau_r=2\text{ms}$ during which the neuron is prevented from firing.

Neurotransmitters were released as a response to a presynaptic action potential fired at time $t_{k}$, binding to the
corresponding receptors at the postsynaptic side of its output neurons. The binding of
neurotransmitters at the receptors triggered the generation of postsynaptic
currents $I^A$ or $I^G$, depending on the presynaptic neuronal type. The total
input current received by a given neuron was described by
\begin{eqnarray}
  I_i^x(t) = g^x\sum_{j=1}^N\sum_{t_j^k}A_{ij}E^x_j(t)\alpha(t-t_j^k-t_d^x),
  \label{eq:syn}
\end{eqnarray}
where $t_d^x$ is a transmission delay (mimicking axonal conduction), with $t_d^A=1.5\text{ms}$ and $t_d^G=4.5\text{ms}$.
$g^x$ is the synaptic strength, which was adjusted to obtain the desired burst
rate. The value of $g^A=7.75\text{pA}$ in a network with inhibition
silenced provided a bursting rate of $\sim$0.1 Hz. When inhibition was active, a comparable bursting rate of of $\sim$0.12 Hz was obtained by setting $g^G=-2g^A$. $E^x_j(t)$ is a term accounting for short--term synaptic depression, and $\alpha(t)$ is an alpha shaped
function of the form
\begin{eqnarray}
  \alpha(t)=\exp\left(1-t/\tau_s\right)\frac{t}{\tau_s}\Theta(t),
  \label{eq:alphashape}
\end{eqnarray}
where $\tau_s=2\text{ms}$ represents the synaptic rise time and $\Theta(t)$ is the Heaviside step function.

Short--term synaptic depression accounts for the depletion of available
neurotransmitters at the presynaptic terminals due to repeated
activity~\cite{Zucker:2002dg}. The neurotransmitters dynamics at the synapses of
neuron $i$ was described by the set of equations~\cite{Tsodyks:1997gu}:
\begin{eqnarray}
  \frac{dR^x_i}{dt}&=&\frac{1-R^x_i-E^x_i}{\tau^x_r}-U\sum_{t_k}R^x_i(t_i^k) \, \delta(t-t_i^k),\nonumber\\
  \frac{dE^x_i}{dt}&=&-\frac{E^x_i}{\tau_i}+U\sum_{t_k}R^x_i(t_i^k) \, \delta(t-t_i^k),
  \label{eq:std}
\end{eqnarray}
where $R^x_i$ and $E^x_i$ are the fraction of available neurotransmitters in the
recovered and active states, respectively. $\tau^x_r$ is the characteristic recovery
time with $\tau^A=_r5000$ ms and $\tau^G_r=100$ ms. $\tau_i=3$ ms is the
inactivation time and $U=0.3$ describes the fraction of activated synaptic
resources after an action potential.

\subsubsection*{Simulating calcium fluorescence signals}
Based on the simulated spike data, synthetic calcium fluorescence signals were generated according to a model that incorporates the calcium dynamics in the neurons and experimental artifacts. The former describes the saturating nature of calcium concentration bound to the calcium dye inside the cells, while the latter treats the noise of the recording camera as well as light scattering due to anisotropies in the recording
medium~\cite{Stetter:2012fe}.

Each action potential of a neuron~$i$ at time~$t$ leads to the intake of ~$n_{i,t}$ calcium ions through the cell membrane, raising the calcium concentration inside the cell. A number $[\text{Ca}^{2+}]_{i,t}$ of the Calcium ions bind the fluorescence dye by a fixed amount $A_{\text{Ca}} = 50
\mu\text{M}$, and are slowly freed with a time scale $\tau_{\text{Ca}} = 1\text{s}$. This process is described by the equation
\begin{equation}
  [\text{Ca}^{2+}]_{i,t} - [\text{Ca}^{2+}]_{i,t-1} = - \frac{\tau_{\text{image}}}{\tau_{\text{Ca}}} [\text{Ca}^{2+}]_{i,t-1} + A_{\text{Ca}} \, n_{i,t},
\end{equation}
where $\tau_{\text{image}}$ is the simulated image acquisition frame rate.

The level of calcium fluorescence~$F^0_{i,t}$ emitted by a cell was modeled by a
Hill function of the bound calcium concentration (with saturation level
$K_d = 300 \mu\text{M}$) together with an additive Gaussian noise term~$\eta_{i,t}$ characterized with a
standard deviation~$\sigma_{\text{noise}} = 0.03$~\cite{Mishchenko:2011ui}, i.e.
\begin{equation}
    F^0_{i,t} = \frac{[\text{Ca}^{2+}]_{i,t}}{[\text{Ca}^{2+}]_{i,t} + K_d} + \eta_{i,t}.
\end{equation}
The level of fluorescence recorded by the camera at a given neuron was \emph{not}
independent of neighboring cells due to the introduction of simulated light scattering. We incorporated this artifact by adding to the monitored cell a fraction $A_{\text{sc}} = 0.15$ of the
fluorescence from neighboring cells, which was weighted according to their mutual
distance~$d_{ij}$ by a Gaussian kernel of width $\lambda_{\text{sc}} =
0.05\text{mm}$. The total fluorescence captured in a neuron was then given by:
\begin{equation}
    F_{i,t} = F^0_{i,t} + A_{\text{sc}} \displaystyle \sum_{j=1, j \neq i}^N F^0_{j,t} \exp \left\{ - \left(d_{ij}/\lambda_{\text{sc}} \right)^2
    \right\}.
\end{equation}

\subsection*{Generalized Transfer Entropy}
Generalized Transfer Entropy~(GTE) was introduced
in~\cite{Stetter:2012fe} as an extension of the original Transfer Entropy notion~\cite{Schreiber:2000jx}. It is given by the Kullback-Leibler
divergence between two probabilistic transition models for a given time series~$I$, conditioned on the system visiting a specified target dynamical state.  In the case of fluorescence signals, this state
selection is achieved by conditioning the analysis to the regime where the
population average of the time series~$G$ is lower than a given
threshold~$\tilde{g}$, i.e.
\begin{equation}
  \text{GTE}_{J \rightarrow I} = \displaystyle \sum P(i_t, i_{t-1}^{(k)}, j_{t-1+S}^{(k)} | \, g_t < \tilde{g}) \log \frac{P(i_t | \, i_{t-1}^{(k)}, j_{t-1+S}^{(k)}, g_t < \tilde{g})}{P(i_t | \, i_{t-1}^{(k)}, g_t < \tilde{g})}.
  \label{eq:GTE}
\end{equation}
Here, vectors in time are denoted by their length in brackets, which is equal to
the order of Markov order approximation assumed for the underlying process, $x_t^{(n)} = \{ x_t, x_{t-1}, ..., x_{t-n+1} \}$.  The sum is defined over all possible values of~$i_t$ and the
vectors $i_{t-1}^{(k)}$ and $j_{t-1+S}^{(k)}$.  The shift variable~$S \in \{0,
1\}$ denotes the inclusion of same-bin (instantaneous) interactions for $S = 1$.
This adjustment was introduced in~\cite{Stetter:2012fe} to cope with the limited
time-resolution of calcium fluorescence signals and is dubbed in the text as
Instantaneous Feedback Term (IFT) correction. Furthermore, the time-series of calcium fluorescence were high-pass filtered by mean of a discrete difference operator, as a straightforward attempt to enhance the visibility of firing events drowned in noise.
Note that GTE reduces to conventional Transfer Entropy for $S = 0$ and
$\tilde{g} \rightarrow \infty$, i.e. when same-bin interactions are excluded and when the
selected state encompasses the whole observed dynamics.
The Markov order of the underlying process is here somewhat arbitrarily set to $k=2$, following on~\cite{Stetter:2012fe} where we extensively checked its effect on the reconstructions: in our previous study, $k=2$ resulted to be the lowest dimensionality in the probability distribution allowing to separate actual interactions from signal artifacts like light scattering. 

Note that we did not perform any delay embedding of the time-series, because we did not find it here necessary to reach satisfying performance levels, or leading to noticeable improvements. Methodological developments along the lines of~\cite{Wibral2011TE-MEG,Vicente:2011ix} would be however desirable for future applications to real experimental data.

Code for our Generalized Transfer Entropy method is publicly available at
https://github.com/olavolav/TE-Causality.

\subsection*{Optimal binning}

The probability distributions in GTE as defined in Eq.~(\ref{eq:GTE}) were
estimated based on discretized values of the temporal difference signal of the
observed fluorescence. To cope with potential undersampling artifacts ---since the probability distributions to estimate have an elevated dimensionality, as large as $2 \, k + 1$--- we symbolized the signals into a binary
sequence, by applying a sharp threshold. The optimal threshold value~$\hat{x}$ for this conversion was obtained from the following analysis. We
first ignored the exponential decay of the
fluorescence signal since it has a small influence on discretely differentiated signals, and
assumed a sufficiently low firing rate so that the occurrence of more than one
spike per frame of a given neuron is negligible.
Under these simplification hypotheses, the probability
distribution of the signal can be cast as a combination of Gaussian functions, with mean values given
by the offset associated to the number of action potentials encountered in the
current time bin. Additionally, to preserve information about spiking events when projecting the time-series into a binary representation, we computed the optimal mapping by
determining the probability $P$ that the mapping is correct at any given time step (provided the
parameters of the model~$\vartheta$ and a threshold value~$x$), i.e.:
\begin{equation}
  P(\text{correct mapping} \,|\, \vartheta) = P(x_t \geq x, s_t = 1 |\, \vartheta) + P(x_t < x, s_t = 0 |\, \vartheta),
  \nonumber
\end{equation}
where~$s_t \in \{0, 1\}$ denotes the occurrence of a firing event at time frame~$t$, and $\vartheta$ refers to unspecified but frozen parameters of the analyzed system, which have a potential influence on the estimated probability. In particular, the probability that a neuron fires at a given image frame is a function of the
firing rate and the length of the image frame, $p_{\text{sp}} = f_{\text{sp}} \,
\tau_{\text{image}}$.  For a normally distributed camera noise with standard
deviation~$\sigma_{\text{noise}}$ and an expected variation~$\Delta x$ in fluorescence due to
a single spike, a straightforward solution for the optimal separation
value~$\hat{x}$ that yields the maximum of the correct mapping probability can be derived:
\begin{equation}
  \hat{x} = \frac{1}{2} \Delta x + \frac{\sigma_{\text{noise}}^2}{\Delta x} \, \log \left(\frac{1-p_{\text{sp}}}{p_{\text{sp}}}\right).
  \label{eq:optimal_binedge}
\end{equation}
GTE scores were robust against the selection of a separation value above the optimal $\hat x$. Indeed, for $x > \hat{x}$ the total number of samples above the separating value is reduced, but the fraction of samples that correspond to real spikes is actually increased. The resulting network reconstructions did not show any notable decrease of quality for values of $x$ up to a 30\% above the optimal value.

\subsection*{Network reconstruction}

In order to reconstruct a whole network, GTE was computed for each
directed pair of neurons~$i,j$ from Eq.~(\ref{eq:GTE}), resulting in a matrix~$M$
of directed causal influences where $M_{ji}=GTE_{J\to I}$. A new binary
matrix~$T(z)$ was created from the GTE scores, where $T_{ji}=1$ if~$M_{ji}$ is amongst the fraction~$z$ of links with the highest GTE score.

The quality of the reconstruction was quantified through a Receiver Operating
Characteristic (ROC) analysis. The ROC is a parametric curve that establishes a
relationship between the true and the false positive links found in~$T(x)$ for
the different thresholded levels. If~$A$ denotes the binary connectivity matrix
of the real network, then the true positive ratio~(TPR) is defined as the number
of links in~$T$ that are present in~$A$ respect to the total number of existing
links. The false positive ratio~(FPR) is the fraction of links in~$T$ that do
not match original links, i.e.,
\begin{eqnarray}
  \text{TPR}(z) &=&
  \displaystyle\sum_{\forall i,j}T_{ji}(z)A_{ji}\Bigg/\displaystyle\sum_{\forall
  i,j}A_{ji},\\ \text{FPR}(z) &=& \displaystyle\sum_{\substack{\forall i,j \\
    i\neq j}}T_{ji}(z)\hat A_{ji}\Bigg/\displaystyle\sum_{\substack{\forall i,j
    \\ i\neq j}}\hat A_{ji},
    \label{eqn:TPR}
  \end{eqnarray}
where $\hat A$ is the negation of the binary connectivity matrix~$A$
($0 \leftrightarrow1 $). Thus $\text{TPR}(z)$ and $\text{FPR}(z)$ constitute, respectively,
finite-size estimates of the probabilities $P(\text{reconstruction} = 1 | \,
\text{true} = 1, z)$ and $P(\text{reconstruction} = 1 | \, \text{true} = 0, z)$,
 for any given link across the network. Confidence intervals for ROC curves were estimated based on 5 different network realizations.

\subsubsection*{Combining two reconstruction results}

To distinguish between excitatory and inhibitory neurons, we combined the
information of the reconstructions obtained from the ``E+I'' and ``E--only'' data, namely $M^{E+I}$ and $M^{Eonly}$. We assumed that excitatory links are present in both datasets, while inhibitory ones appear only in the ``E+I'' reconstruction, and proceeded by defining new matrices of putative excitatory~$M^{\text{exc}}$ and putative inhibitory influences~$M^{\text{inh}}$, of the form:
\begin{eqnarray}
  M^{\text{exc}} = M^{E+I} + M^{Eonly}, \\
  M^{\text{inh}} = M^{E+I} - M^{Eonly}.
\end{eqnarray}
To obtain the effective connectivity reconstruction only the rank ordering
of GTE values is relevant. Therefore no rescaling of these matrices is necessary, and the final set of links could be obtained by thresholding the matrices as described above.

To label the neurons as either excitatory or inhibitory, we first
removed all links that were present in both reconstructions, and then ranked
the neurons according to the difference between excitatory and inhibitory links, $l_i
= \sum_j T^{\text{exc}}_{ji} - \sum_j T^{\text{inh}}_{ji}$.  We next used the
prior information that a fraction~$f_E = 80$\%
of the neuronal population is excitatory, therefore identifying as excitatory neurons the~$f_E$ fraction with the highest $l_i$~score, and labeling the rest as inhibitory.

\subsubsection*{Statistical tests}

Statistical
significance on the inference of excitatory and inhibitory neuronal types was
performed as follows. Assuming that the fraction of excitatory and inhibitory
neurons ($f_E$ and $f_I$ respectively) is known with good precision in a population of $N$ cells, the
probability to correctly identify by chance a given set of neurons $n_E$ and
$n_I$ in a given trial $X$ follows a binomial distribution:
\begin{eqnarray}
  P(X=n) = \binom{Nf_x}{n}f_x^{n}(1-f_x)^{Nf_x-n}.
  \label{eq:binomial:}
\end{eqnarray}
Let suppose that a labeling method provides a fraction $n_{\textrm{guess}}$ of correctly labeled links. We considered this labeling result as statistically significant if the probability of outperforming by chance this success rate was $P(X\geq n_\textrm{guess}) < p$, with a standard choice of  $p=0.05$.

\section*{Acknowledgments}
JO and JS received financial support from Ministerio de Ciencia e Innovaci\'on
(Spain) under projects FIS2009-07523, and FIS2010-21924-C02-02, FIS2011-28820-C02-01 and from the Generalitat de Catalunya under project 2009-SGR-00014. OS, TG and DB were supported by the German Ministry for Education and Science~(BMBF) via the Bernstein Center for Computational Neuroscience~(BCCN) G\"ottingen (Grant No.~01GQ0430). OS and DB acknowledge in addition, respectively, the Minerva Foundation (M\"unchen, Germany) and funding from the FP7 Marie Curie career development fellowship IEF 330792 (DynViB).

\bibliography{Orlandi2014-Revision_3}
\newpage
\section*{Figure Legends}

\end{document}